%% file: prd_v6_0.tex
\documentstyle[preprint,eqsecnum,epsfig,aps,floats,tighten]{revtex} % for preprint mode
\def\pbar{$\overline{p}~$}               %pbar
\def\tbar{$\overline{t}~$}               %tbar
\def\ppbar{$p\overline{p}~$}             %ppbar
\def\ttbar{$t\overline{t}~$}             %ttbar
\def\et{$E_T$}                          %ET
\def\met{\mbox{${\hbox{$E$\kern-0.6em\lower-.1ex\hbox{/}}}_T$}} %missing ET
\def\ptw{$p_T^W$} 
\def\D0{D\O}                            %D0
\newcommand{\nunit}[2]{#1\,\mbox{#2}}
\begin{document}
\lefthyphenmin=2
\righthyphenmin=3

%new 060600
%\setlength{\baselineskip}{24pt}
%\preprint{FERMILAB-Pub-97/109-E}
%
% ======> Title of the paper goes here <====================
%
\title{
Measurement of the Angular Distribution of Electrons from $W
\to e \nu$ Decays Observed in $p \bar p$ Collisions at $\sqrt{s}= 1.8$~TeV\\
}

\input{list_of_authors.tex}

\maketitle
%
% ==============> Text of the abstract goes here <=====================
% 
\begin{abstract}
We present the first measurement of the electron angular
distribution parameter ${\alpha_{2}}$
in $W \to e \nu$ events produced in proton-antiproton collisions 
as a function of the $W$ boson transverse momentum. 
Our analysis is based on data collected using the \D0 
detector during the 1994--1995 Fermilab Tevatron run. 
We compare our results with
next-to-leading order perturbative QCD, which predicts an angular
distribution of ($1 \pm \alpha_1 \, \mathrm{cos} \theta^* +
\alpha_2\,  \mathrm{cos}^2 \theta^*$), where $\theta^*$ is the
polar angle of the electron in the
Collins-Soper frame.  In the presence of QCD corrections, the
parameters $\alpha_1$ and $\alpha_2$ become functions of $p_T^W$, the
$W$ boson transverse momentum.  
This measurement 
provides a test of next-to-leading order QCD
corrections which are a non-negligible contribution to the $W$ boson mass
measurement.
\end{abstract}
%\pacs{PACS numbers 14.65.Ha, 13.85.Qk, 13.85.Ni}
\newpage
\vfill\eject
%\twocolumn
\section{Introduction}
\label{sec:intro}
After the discovery of the $W$ boson~\cite{ua1disc,ua2disc}
at the CERN \ppbar\ collider, 
early studies of its properties verified its left-handed coupling to  
fermions and established it to be a spin~1 particle~\cite{ua1cos,ua1}.
These were accomplished through the  measurement of 
the angular distribution of the 
charged lepton from the $W$ boson decay, a measurement ideally suited
to \ppbar colliders. The angular distribution was found to follow 
the well-known $V-A$ form $(1\pm\cos\theta^*)^2$,
where the polar angle 
$\theta^*$ is the lepton direction in the rest frame of the $W$ boson 
relative to the 
proton direction, and the sign is opposite that of the charge of the 
$W$ boson or emitted lepton; 
this formulation assumes that only valence quarks participate 
in the 
interaction, otherwise the angular distribution is slightly modified. It is
important to note that 
these measurements were performed on $W$ bosons produced with almost no
transverse
momenta. This kinematic region is dominated by the production 
mechanism $\bar{q}+q'\rightarrow W$. The center of mass energy used, $\sqrt{s}
=\nunit{540}{GeV}$, is not high enough for other processes to 
contribute substantially.

\begin{figure}[!htbp]
%\vspace{-1in}
\centerline{\epsfig{file=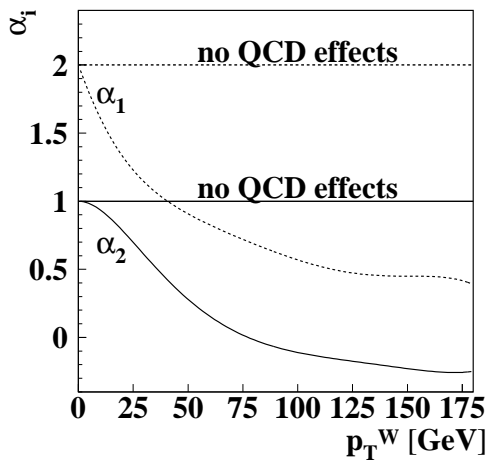,width=7cm}}
%\vspace{-1.8in}
\caption{The angular parameters $\alpha_1$ (dashed) and $\alpha_2$ (solid) 
as functions of \ptw. These parameters are evaluated integrated over 
the $W$ boson rapidity, $y$. In the absence of QCD effects $\alpha_1$ and 
$\alpha_2$ equal 2.0 and 1.0, respectively.\label{fig:alpha12}}
\end{figure}

At the higher energies of the Fermilab Tevatron ($\sqrt{s}=\nunit{1.8}{TeV}$) 
and higher transverse momenta explored 
using the \D0 detector~\cite{detector}, other processes
are kinematically allowed to occur. At low $W$ boson transverse momentum, 
$p_T^W$, 
the dominant higher order process involves 
initial state radiation  of soft gluons. 
This process is calculated through the use
of resummation techniques as discussed in Refs. 
\cite{csa,davstira,altarellis,davstirb,ly,AKtheory,resb}.
At higher values of $p_T^W$, where perturbation theory holds, 
other processes contribute~\cite{ARtheory}, such as:
\begin{enumerate}
  \item $\bar{q}+q'\rightarrow W + g$
  \item $q+g\rightarrow W + q'$
  \item $g+g\rightarrow W + \bar{q}+q'$
\end{enumerate}
where only the first two contribute significantly at Tevatron
energies \cite{mi}.
These two processes change the form of the angular distribution of the
emitted charged lepton to 
\begin{equation}
  \frac{d\sigma}{dp^2_T\,dy\,d\!\cos\theta^*}\propto(1\pm
                 \alpha_1 \cos\theta^* + \alpha_2 \cos^2\theta^*)
                 \label{eq:angdistth}
\end{equation}
where
the parameters $\alpha_1$ and $\alpha_2$ depend on the $W$ boson $p_T$ 
and rapidity, $y$~\cite{mi}. 
In Fig.~\ref{fig:alpha12}, the parameters  $\alpha_1$ and $\alpha_2$ 
are shown as functions of \ptw. 
The angle $\theta^*$ is measured in the Collins-Soper 
frame \cite{cs}; this is the rest frame of the $W$ boson where the $z$-axis
bisects the angle formed by the proton momentum and the negative of the
antiproton
momentum with the $x$-axis along the direction of $p_T^W$. 
This frame is chosen since it reduces the ambiguity of the 
neutrino longitudinal momentum to a sign ambiguity on $\cos\theta^*$.

In this paper, we present the first measurement of $\alpha_2$ as a function of $p_T^W$~\cite{gsthesis},
which serves as a probe of next-to-leading order quantum chromodynamics 
(NLO QCD), using the well-understood 
coupling between $W$ bosons and fermions. 
This measurement probes the effect of QCD corrections on 
the spin structure of $W$ boson production.

At \D0, the most precise $W$ boson mass measurement is made by fitting 
the transverse mass distribution. However, since the transverse mass of 
the $W$ boson
is correlated with the decay
angle of the lepton, the QCD effects discussed above introduce a 
systematic shift $\sim 40$ MeV 
to the $W$ boson mass measurement 
for events with $p_T^W\leq $ 15~GeV 
which must be taken into account.
Presently, the Monte Carlo program used in the mass measurement 
models the
angular distribution of the decay electron 
using the calculation of Mirkes~\cite{mi}. 
During the next run of the Fermilab Tevatron collider (Run II), 
when the total error on the $W$ boson mass
will be reduced from the current 91 MeV for \D0
~\cite{run1awmassprl,run1awmassprd,wmassprd,wmassprl,wmassprd_ec,wmassprl_ec} 
to 
an estimated 50 MeV for 1 $\rm fb^{-1}$ and  to about 
30 MeV for 10 ${\rm fb}^{-1}$~\cite{TEV2000}, a good understanding of
this
systematic shift is important. Therefore,
a direct measurement of 
the electron angular decay distribution is important to minimize
the systematic error.

The paper is organized as follows: a brief description of the 
\D0 detector is given in Sec.~\ref{sec:det}, with an emphasis on the components
used in this analysis. 
Event selection is discussed in Sec.~\ref{partid}. 
The analysis procedure is described in Sec.~\ref{sec:analy}. Finally,
conclusions are presented in Sec.~\ref{sec:concl}.
\section{The \D0 Detector}
\label{sec:det}
\subsection{Experimental Apparatus}
The \D0 detector, described in more detail elsewhere~\cite{detector}, is
composed of four major systems. The innermost of these is a non-magnetic
tracker used in the reconstruction of charged particle 
tracks. The tracker is surrounded by  central and forward 
uranium/liquid-argon sampling calorimeters. 
These calorimeters are used to identify
electrons, photons, and hadronic jets, and to reconstruct their energies.
The calorimeters are surrounded by a muon spectrometer which is composed of 
an iron-core toroidal magnet surrounded by drift tube chambers. The system is
used in the identification of muons
and the reconstruction of their momenta. 
To detect inelastic $p$\pbar\ collisions for triggering, 
and to measure the luminosity, a set of scintillation counters
is located in front of the forward calorimeters.
For this analysis, the relevant
components are the 
tracking system and the calorimeters.
We use a coordinate system where the polar angle $\theta$ is measured 
relative to the proton beam direction $z$, and $\phi$ is the azimuthal angle.
The pseudorapidity $\eta$ is defined as $-\ln(\tan\frac{\theta}{2})$, and 
$\rho$ is the perpendicular distance from the beam line.

The structure of the calorimeter has been optimized to distinguish 
electrons and photons 
from hadrons, and to measure their energies. 
It is composed of three
sections: the
central calorimeter (CC), 
and two end calorimeters (EC).
The $\eta$-coverage for electrons used in this analysis 
is $\mid \eta\mid<1.1$ in the CC
and $1.5<\mid \eta\mid<2.5$ for the EC.
The calorimeter is segmented
longitudinally into two sections, the
electromagnetic (EM) and the hadronic (HAD) calorimeters. The primary
energy
measurement needed in this analysis comes from the EM calorimeter,
which is subdivided longitudinally into four layers (EM1--EM4). 
The hadronic calorimeter is subdivided longitudinally into four  
fine hadronic layers (FH1--FH4) and one course hadronic layer (CH).  
The first, second and
fourth layers of the EM calorimeter are transversely divided into  
cells of size $\Delta\eta\times \Delta\phi = 0.1 \times 0.1 $.
The shower 
maximum occurs in the third layer, which is divided into 
finer units of $ 0.05 \times 0.05 $ to improve the shower 
shape measurement.
\subsection{Trigger}
\label{sec:trig}
The \D0 trigger is built of three levels, with each level applying
increasingly more sophisticated selection criteria on an event. The
lowest
level trigger, Level 0, uses the scintillation counters in front of the
forward calorimeters to signal the presence of an inelastic \ppbar\
collision.
Data from the Level 0 counters, the calorimeter and the muon chambers 
are sent to the Level 1 trigger,
which allows the experiment to be triggered on 
total transverse energy, \et, missing transverse energy, \met, 
\et\ of individual calorimeter towers, and/or the presence
of a muon. 
These triggers operate in less than  $3.5\;\mu\mbox{s}$, the
time between bunch crossings.
A few calorimeter and muon triggers require additional
time, 
which is provided by a Level 1.5 trigger system.

Candidate Level 1 (and 1.5) triggers initiate the Level 2 trigger
system that 
consists of a farm of microprocessors. These microprocessors run
pared-down versions of the off-line analysis code to select events based
on
physics requirements. Therefore, the experiment can be triggered on
events
that have characteristics of $W$ bosons or other physics criteria.

\section {Particle Identification and Data Selection}
\label{partid}
This analysis relies on the \D0 detector's ability to identify 
electrons and the undetected energy associated with 
neutrinos. The particle identification techniques employed are 
described in greater detail in Ref.~\cite{wzcsprd}. The following 
sections provide a brief summary of the techniques used in this paper.
\subsection{Electron Identification} 
Identification of electrons starts at the trigger level, where clusters
of electromagnetic energy are selected. At Level 1, the trigger searches for
EM calorimeter towers ($\Delta\phi\times\Delta\eta=0.1\times0.1$) that exceed
predefined thresholds. $W$ boson triggers require 
that the energy deposited in a
single EM calorimeter tower exceed \nunit{10}{GeV}.
Those events that satisfy the Level 1 trigger
are processed by the Level 2 filter. The trigger towers are combined 
with energy in the surrounding calorimeter cells within a window of
$\Delta\phi\times\Delta\eta=0.3\times0.3$. Events are selected 
at Level 2 if 
the transverse energy in this window exceeds 20 GeV.
In addition to the $E_T$
requirement, the longitudinal and transverse shower shapes 
are required to match
those expected for electromagnetic showers.
The longitudinal shower shape is described by  
the fraction of the energy deposited in each of the four EM layers of the 
calorimeter. The transverse shower shape is characterized by the 
energy deposition patterns in the third EM layer. The difference 
between the energies in concentric regions covering
$0.25\times 0.25$ and $0.15\times 0.15$ in $\eta\times\phi$ must be 
consistent with that expected for an electron~\cite{detector}.

In addition, at Level 2, the energy cluster isolation
is required
to satisfy $f_{iso}<0.15$, where $f_{iso}$ is defined as:
\begin{equation}
  f_{iso}=\frac{E_{{\rm total}}(0.4)-E_{{\rm EM}}(0.2)}
                      {E_{{\rm EM}}(0.2)},
\end{equation}
$ E_{{\rm total}}(0.4)$ is the total energy, and
$E_{{\rm EM}}(0.2)$ the electromagnetic energy, in cones of 
$R=\sqrt{(\Delta\eta)^2+(\Delta\phi)^2} =0.4$ and $0.2$, 
respectively.
This cut preferentially selects the isolated electrons expected from vector 
boson decay. 

Having selected events with isolated electromagnetic showers at the trigger
level, a set of tighter cuts is imposed off-line to identify
electrons, thereby reducing the background from QCD multijet events. 
The first step
in identifying an electron is to build a cluster about the trigger tower using
a nearest neighbor algorithm. As at the trigger level, the cluster is required
to be isolated ($f_{iso}<0.15$). To increase the likelihood
that the cluster is due to an 
electron and not a photon, a track from the central tracking system is 
required to point at its centroid.
We extrapolate the track to the third EM layer in the calorimeter and 
calculate the distance between the extrapolated track and the cluster 
centroid in the azimuthal direction, $\rho \Delta \phi$, and in the 
$z$-direction, $\Delta z$. The cluster centroid position is extracted at 
the radius of the third EM layer of the calorimeter, $\rho$.
%
%electron revertexing here
The $z$ position of the event vertex is defined by the line connecting the 
center of gravity calorimeter position of the electron and the center of 
gravity of its associated track in the central tracking system, 
extrapolated to the beamline. The electron \et\ is 
calculated using this vertex definition~\cite{wzcsprd}.
The variable 
\begin{equation}
\sigma_{trk}^2=\left( \frac{\rho\Delta \phi}{\sigma_{\rho \phi}}\right)^2 +
\left( \frac{\Delta z}{\sigma_z}\right)^2 \label{eq:trk_pos}
\end{equation}
where  $\sigma_{\rho \phi}$ and $\sigma_z$ are the respective track 
resolutions, 
quantifies the quality of the match. 
A cut of $\sigma_{trk} < 5$ is imposed on the data.
Electromagnetic clusters that satisfy these criteria, referred to as 
``loose electrons,'' are then subjected to a
4-variable likelihood test previously used in the measurement of 
the top quark mass by the \D0 collaboration~\cite{topmassprd}. 
The four variables are:
\begin{itemize}
  \item A $\chi^2$ comparison of the shower shape with the expected shape 
	of an electromagnetic shower, computed using a $41$-variable 
 	 covariance matrix~\cite{covprd} of the energy depositions in the 
	cells of the 
	electromagnetic calorimeter and the event vertex.

  \item The electromagnetic energy fraction, which is defined as the 
        ratio of shower energy in the EM section of the calorimeter 
	to the total EM energy plus the energy in the first hadronic 
	section of the calorimeter.

  \item A comparison of track position to cluster centroid position as
        defined in Eq.~\ref{eq:trk_pos}.

  \item The ionization, $dE/dx$, along the track, to reduce  
	contamination from  $e^+e^-$ pairs due to photon conversions. 
	This variable is effective in reducing the background from 
	jets fragmenting into neutral pions which then decay into photon pairs.

\end{itemize}
To a good approximation, these four variables are independent of each other 
for electron showers. 
Electrons that satisfy this additional cut are called ``tight'' electrons.

\subsection{Missing Energy} 
The primary sources of missing energy in an event include the neutrinos 
that pass through
the calorimeter undetected and  the apparent
energy imbalance due to calorimeter  resolution. 
The energy imbalance is measured only in the
transverse plane 
due to the unknown momenta of the particles 
escaping within the beam pipes. 

The missing transverse energy is calculated by taking the
negative of the vector sum of the transverse energy in all of the 
calorimeter cells. 
This gives both the
magnitude and direction of the \met, allowing the 
calculation of the
transverse mass of the $W$ boson candidates, $M_T^W$, 
given by 
\begin{equation}
M_T^W = \sqrt{2E_T^e \met[1-\cos(\phi^e-\phi^\nu)]}
\end{equation}
in which $E_T^e$ is the transverse energy of the electron 
and $\phi^e$ and  $\phi^\nu$
are the azimuthal angles of the electron and neutrino, respectively.
\subsection{Event Selection}
The $W$ boson data sample used in this analysis
was collected during the 1994--1995 run of the Fermilab
Tevatron collider. This data sample corresponds to an integrated luminosity of 
$85.0\pm 3.6\;\rm pb^{-1}$.
Events are selected by 
requiring one tight electron in the central 
calorimeter $(\mid \eta \mid <1.1)$ with \et\ $>25$ GeV.
The CC consists of 32 $\phi$ modules.
To avoid areas of reduced response between neighboring 
modules, the $\phi$ of an electron is 
required to be at least $0.05\times2\pi/32$ radians away from the 
position of a module boundary. 
In addition, events are required to have \met\ $>25$ GeV.
If there is a second electron in the event (loose or tight) and the 
dielectron invariant mass $M_{ee}$ is close to the $Z$ boson mass 
($75 \;\mbox{GeV} < M_{ee} < 105 \;\mbox{GeV}$), the event is rejected.

To ensure a well-understood 
calorimeter response and to reduce luminosity-dependent effects,
two additional requirements are imposed.
The Main Ring component of the Tevatron accelerator passes 
through the  outer part of the hadronic calorimeter. Beam losses 
from the Main Ring can cause significant energy deposits in the 
calorimeter, resulting in false \met. The largest losses occur when beam is 
injected into the Main Ring. Events occurring within a 400 ms 
window after injection are rejected, resulting in a 17\% loss of data.
Large beam losses can also 
occur when particles in the Main 
Ring pass through the \D0 detector. Hence we reject events within a 
1.6 $\mu\rm{s}$  window around these occurrences, 
resulting in a data loss of approximately
$8\%$. 
After applying all of the described cuts, a total of 
41173 $W$ boson candidates is
selected using electrons found in the central calorimeter.
\section{Experimental Method}
\label{sec:analy}
\subsection {Monte Carlo Simulation \label{sec:mc}}
For  this analysis, 
a Monte Carlo program with a parameterized detector simulation is 
used. This is the same Monte Carlo used in our previous results on
the $W$ boson mass measurement~\cite{wmassprd} and the inclusive cross 
sections of the
$W$ and $Z$ bosons~\cite{wzcsprd}, so it will only be briefly summarized
here.

In the Monte Carlo, the detector response is parameterized using the data
from the experiment. This includes using $Z$ bosons and their hadronic recoil
to study the response and resolution.
The response itself is then parameterized as a function of
energy and angle.

The kinematic variables for each $W$ boson are generated using the 
{\sc resbos}~\cite{resb} event generator  
with the theoretical model described in 
Refs.~\cite{ly,ARtheory},
and the CTEQ4M parton distribution functions (pdf's)~\cite{cteq4}. 
Finally, the
angular distribution is generated according to the calculation
of Mirkes~\cite{mi}.
\subsubsection{Hadronic Scale\label{sec:hadresponse}} 
One of the parameters needed for the Monte Carlo program used in this study
is the response of the calorimeter 
to the hadronic recoil, defined as the sum of all calorimeter cells 
excluding the cells belonging to the electron.
The detector response and resolution for particles recoiling against a 
$W$ boson should be the same as for particles recoiling against a $Z$ 
boson. For $Z\rightarrow ee$ events, we measure the transverse momentum 
of the $Z$ boson from the $e^+e^-$ pair, $p_T^{~ee}$,
and from the recoil jet momentum, $p_T^{~\rm rec}$, in the same manner as 
for $W\rightarrow e\nu$
events. By comparing $p_T^{~ee}$ and $p_T^{~\rm rec}$,
the recoil response is calibrated relative to the well-understood electron 
response~\cite{wmassprd}.

The recoil momentum is carried by many particles,
mostly hadrons, with a wide momentum spectrum. 
Since the response of calorimeters to hadrons tends to be 
non-linear and the recoil particles are distributed over 
the entire calorimeter, including module boundaries with reduced response, 
we expect a momentum-dependent response function with values below unity. 

\begin{figure}[!htbp]
\vspace{-0.7in}
\centerline{\epsfig{file=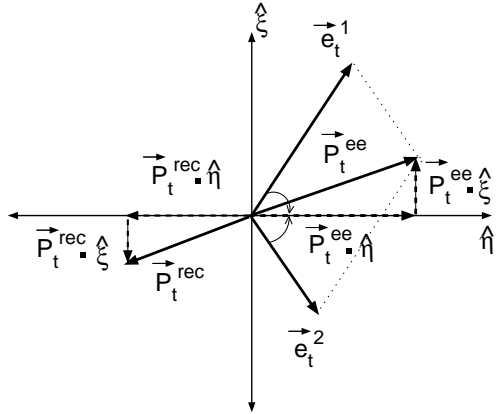,width=12cm}}
\vspace{-1.8in}
\caption[Definition of the $\eta$--$\xi$ coordinate system in a $Z$
boson event.] {Definition of the $\eta$--$\xi$ coordinate
system in a $Z$ boson event. 
$\vec{e}_t^{~i}$ denote the transverse momentum vectors of the two electrons.
The $\eta$ axis is the bisector
of the electrons in the transverse plane; the $\xi$ axis is
perpendicular to $\eta$ \cite{wmassprd}.\label{fig:eta-xi-axis}}
\end{figure}

To measure the recoil response from our data, we use 
a sample of $Z$ boson events with one electron in the CC and the second 
in the CC or the EC (CC/CC+EC). This allows 
the rapidity distribution of the $Z$ bosons to approximate that 
of the $W$ bosons where the neutrinos could be anywhere in the detector.
Further, we require that both electrons satisfy the tight 
electron criteria. This reduces the background for the topology 
where one electron is in the EC. 
We project the transverse momenta of the recoil and the 
$Z$ boson onto the inner bisector of the electron directions 
($\eta$-axis), as shown in Fig.~\ref{fig:eta-xi-axis}. By projecting the 
momenta onto an axis that is independent of any energy measurement, 
noise contributions to the momenta average to zero and do not bias the 
result. 

\begin{figure}[!htbp]
\centerline{\psfig{figure=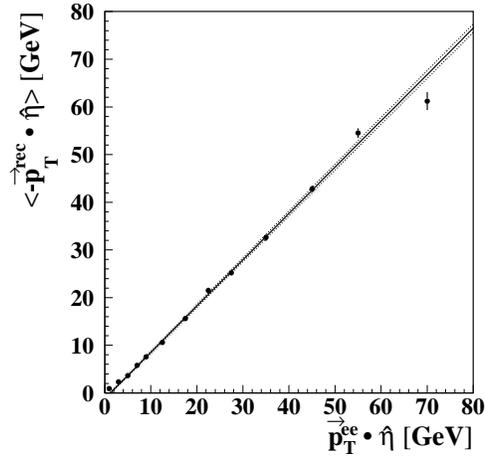,width=7cm}}
\caption{For $Z\rightarrow ee$ events (points) the average value of
$\vec{p}_T^{~\rm rec}\cdot(-\hat{\eta})$ is shown versus $\vec{p}_T^{~ee}\cdot\hat{\eta}$. The line shown is obtained from a linear least squares fit to the
data above $p_T^{ee}$ = 5~GeV as described in the text. 
The dotted lines represent the statistical uncertainties from the fit.\label{HADRONIC_ALL}}
\end{figure}

\begin{figure}[!ht]
\centerline{\epsfig{figure=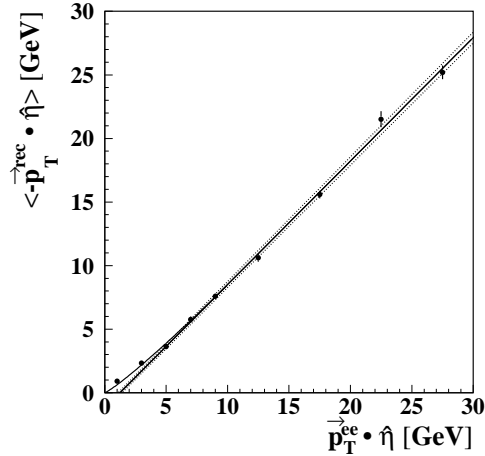,width=7cm}}
\caption{For $Z\rightarrow ee$ events (points) the average value of
$\vec{p}_T^{~\rm rec}\cdot(-\hat{\eta})$ is shown versus 
$\vec{p}_T^{~ee}\cdot\hat{\eta}$. 
Shown is the linear fit valid at $p_T^{ee} > $ 10 GeV and a 
logarithmic fit valid for $p_T^{ee} < 10 $ GeV.
The dotted lines represent the statistical uncertainties from the linear 
fit.\label{HADRONIC}}
\end{figure}

To determine  the functional dependence of the recoil system with respect 
to the dielectron system, $\vec{p}_T^{~\rm rec}\cdot(-\hat{\eta})$ is 
plotted as a function of
$\vec{p}_T^{~ee}\cdot\hat{\eta}$ as shown in Fig.~\ref{HADRONIC_ALL}. 
For $p_T^{ee} > 10 $ GeV, the hadronic response is well described by 
a linear scale  and offset:
\begin{equation}
\vec{p}_{T}^{~\rm rec}\cdot \hat{\eta}   =\alpha_H \; \vec{p}_{T}^{~ee}\cdot\hat{\eta}  +\beta_H 
\end{equation}
The parameters $\alpha_H$ and $\beta_H$ are calculated using
a least-squares fit to the data in the region ${p}_T^{ee}> 5\; {\rm GeV}$, resulting in 
$\alpha_H = 0.972 \pm 0.0095$ and  $\beta_H = (-1.21 \pm 0.14)\; {\rm GeV}$.
For small values of $p_T^{ee}$, $p_T^{ee} < 10$ GeV, the relation between 
the hadronic and electronic recoil 
is best described by a logarithmic function~\cite{wmassprd,jetenim}:
\begin{equation}
\vec{p}_{T}^{~\rm rec}\cdot\hat{\eta}  =\left( \gamma_H \ln ( \vec{p}_{T}^{~ee}\cdot\hat{\eta}) + \delta_H   \right)
\vec{p}_{T}^{~ee}\cdot\hat{\eta} 
\end{equation}
The parameters  $\gamma_H$ and $\delta_H$ are derived using
a least-squares fit to the data 
in the region $p_T^{ee} < 10 \; {\rm GeV}$ (see Fig.~\ref{HADRONIC}), yielding 
$\gamma_H = 0.099 \pm 0.019$ and $\delta_H = 0.620 \pm 0.047 $. In the 
intermediate region, $5\; {\rm GeV} < p_T^{ee} < 10\; {\rm GeV}$, 
the logarithmic and the linear fit match.

\subsubsection {Tuning the Recoil Resolution Parameters}
In the Monte Carlo, we parameterize the calorimeter 
resolution, $\sigma_{\rm rec}$, 
for the hard component of the recoil as
\begin{equation}
\sigma_{\rm rec}=s_{\rm rec}\sqrt{p_T^{~\rm rec}}
\label{eqn:srec}
\end{equation}
where $s_{\rm rec}$ is a tunable parameter, 
and $p_T^{~\rm rec}$ is the recoil momentum 
of the hard component.

The soft component of the recoil is modeled by the 
transverse momentum imbalance from minimum bias events\footnote{Minimum 
bias events are taken with a special 
trigger requiring only that a $p$\pbar\ interaction has taken place. 
The kinematic properties of these events are independent of specific 
hard scattering processes and  model detector resolution effects and pile-up 
which lead to finite \met.}.
This automatically models detector 
resolution and pile-up.
To account for any possible difference between the underlying event in 
$W$ boson events and minimum bias events, we multiply the minimum bias \met\ 
by a correction factor $\alpha_{\rm mb}$. 
We tune the two parameters 
$s_{\rm rec} $ and $\alpha_{\rm mb}$ by comparing the width 
of the $\eta$-balance, 
$\vec{p}_T^{~\rm rec}\cdot \hat{\eta}/R_{\rm rec}+
\vec{p}_{T}^{~ee}\cdot\hat{\eta}$, 
measured from  the CC/CC+EC $Z$ boson data sample 
to Monte Carlo and adjusting the parameters in the Monte Carlo 
simultaneously until the 
widths agree. 
The width of the $\eta$-balance 
is a measure of the recoil momentum resolution.
The recoil response, $R_{\rm rec}$, is defined as
\begin{equation}
R_{\rm rec} = \frac
{|\vec{p}_T^{~\rm rec} \cdot \hat{q}_T|}
{|q_T|},
\end{equation}
where $q_T$ is the generated transverse momentum of the $Z$ boson.
The contribution of the electron momentum resolution to the 
width of the $\eta$-balance is negligibly small. The contribution 
of the recoil momentum resolution grows 
with $\vec{p}_T^{~ee}\cdot\hat{\eta}$ while 
the contribution from the minimum bias \met\ is independent of 
$\vec{p}_T^{~ee}\cdot\hat{\eta}$.
This allows us to determine $s_{{\rm rec}}$ and $\alpha_{\rm mb}$
simultaneously and without sensitivity to the electron resolution 
by comparing the width of the $\eta$-balance predicted by the Monte Carlo
model with that observed in the data in bins of 
$\vec{p}_T^{~ee}\cdot\hat{\eta}$.
We perform a $\chi^2$  fit comparing Monte Carlo and collider data.
The values that minimize the $\chi^2$ are found to be 
$s_{{\rm rec}}=0.665 \pm 0.062\;{\rm GeV^{1/2}} $ 
and  $\alpha_{{\rm mb}}=1.095 \pm 0.020$. 
The non-linear hadronic scale in the region $p_T < $10 GeV leads 
to $s_{{\rm rec}} = 0.50 \pm 0.06 \; {\rm GeV^{1/2}} $, while 
$\alpha_{{\rm mb}}$ is 
unchanged.

\subsection{Extraction of  the Lepton Angle \label{extraction}}
Since only the transverse components of 
the neutrino momentum are measured, the transformation from the lab frame 
to the $W$ boson rest frame (Collins-Soper frame) is not directly calculable.
Therefore
the polar angle of the electron from the $W$ boson decay, $\theta^*$, is not
directly measurable.  In this analysis, $\theta^*$ is inferred from the 
correlation between the transverse
mass of the $W$ boson 
and $\cos\theta^*$ through the use of Bayes' Theorem\cite{jaynes}.

Experimentally, the only information we have about the $W$ boson is that 
contained in the two
kinematic variables $M_T^W$ and $p_T^W$. But 
$M_T^W$ depends on the polar angle $\cos\theta^*$,
the azimuthal angle $\phi^*$ over which we have integrated, and 
$p_T^W$. Therefore, the two experimentally measured
variables $M_T^W$ and $p_T^W$ give $\cos\theta^*$. An analytic
expression exists for this relation (see Ref.~\cite{ma}), 
so in principle the equation is solvable
for $\cos\theta^*$, but the experimental values of 
both $M_T^W$ and $p_T^W$ include detector resolution effects 
that have to 
be unfolded to give the true $\cos\theta^*$ distribution.
Even with perfect detector resolution, the equation would only be solvable 
if the $W$ boson mass was known on an event by event basis. 
Therefore, we calculate 
the probability of measuring $M_T^W$ for a given value $\cos\theta^*$ in
a given $p_T^W$ bin, $p(M_T^W|\cos\theta^*,p_T^W)$. This probability function 
is inverted to 
give the probability of measuring $\cos\theta^*$ for 
a measured $M_T^W$, $p(\cos\theta^*|M_T^W,p_T^W)$, using Bayes' Theorem:
\begin{equation}
  p(\cos\theta^*|M_T^W,p_T^W) = \frac{p(M_T^W| \cos\theta ^{*},p_T^W)p(\cos\theta ^{*})}
        {\int p(M_{T}^W| \cos\theta ^{*},p_T^W)p(\cos\theta ^{*}) d \cos\theta ^{*}} 
\end{equation}
where $p(\cos\theta^*)$ is the prior probability function, which we take as 
$p(\cos\theta^*)=(1+\cos^2\theta^*)$, the charge-averaged expectation from 
$V-A$ theory without QCD corrections.

\begin{figure}[!htbp]
\vbox{
\centerline{
\epsfig{figure=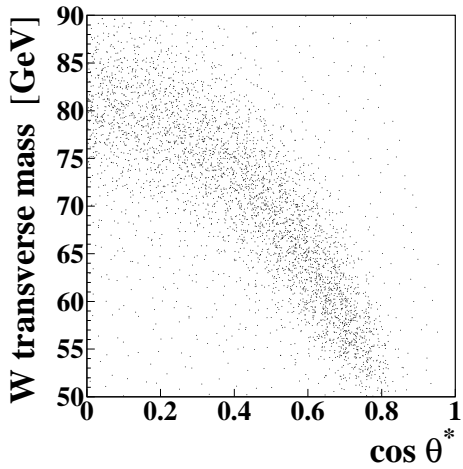,width=7.0cm}}
\caption{Smeared $W$ boson transverse mass versus true $\cos\theta ^*$ for 
         $p_{T}^W\leq$  
10 GeV from Monte Carlo. 
Acceptance cuts have been applied to events in this plot. 
This correlation plot is used to infer the $\cos \theta^*$ distribution 
from the measured $M_T^W$ distribution.}
\label{fig:corr}
}
\end{figure}

To derive the probability function $p(M_{T}^W| \cos\theta ^{*},p_T^W)$, we use
a Monte Carlo simulation of the \D0 detector,
which is described in Sec.~\ref{sec:mc}. 
The correlation between $M_T^W$ and 
$\cos\theta^*$ for $p_T^W\le \nunit{10}{GeV}$ is shown in
Fig.~\ref{fig:corr}.  
After determining
$p(M_{T}^W| \cos\theta ^{*},p_T^W)$, it is inverted, yielding 
$p(\cos\theta^*|M_T^W,p_T^W)$. 
The angular distribution
is calculated by multiplying $p(\cos\theta^*|M_T^W,p_T^W)$ with the measured 
transverse mass distribution. 
This is done in four $p_T^W$ bins covering 0--10 GeV, 10--20 GeV, 
20--35 GeV, and 35--200 GeV.

With the unfolded angular distributions now calculated, the value of
$\alpha_2$ in each of the four $p_T^W$ bins can be determined. This
is accomplished by generating a set of angular distribution templates
for different values of $\alpha_2$. These templates are generated in
a series of Monte Carlo experiments using the Monte Carlo program 
described in Sec.~\ref{sec:mc}.
%\D0 detector simulation 
%with the $W$ bosons being generated using the theoretical model described in 
%Refs.~\cite{ly,ARtheory},
%and the CTEQ4M parton distribution functions (pdf's)~\cite{cteq4}. 

\begin{figure}[!htbp]
\vbox{
%\vspace{-0.5in}
\centerline{
\epsfig{figure=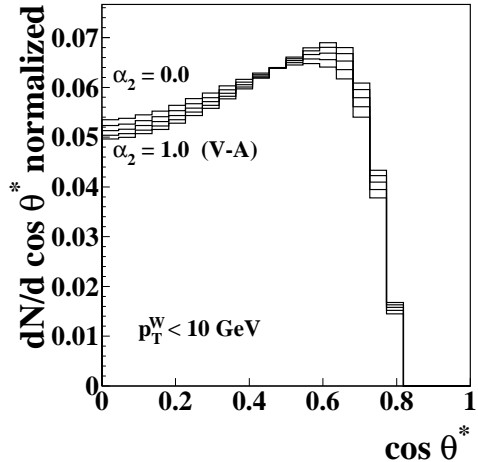,height=7cm,width=7cm}}
%\vspace{-0.5cm}
\caption{Templates of the angular distribution for various $\alpha_{2}$ 
values for $p_T^W \leq$ 10 GeV. These templates are obtained from Monte
Carlo after acceptance cuts have been applied which results in the drop-off at 
small angles. Each template is normalized to unity.}
\label{fig:temp}
}
\end{figure}

The $\cos\theta^*$ templates are 
compared to the data through the use of a maximum likelihood method. 
Fig.~\ref{fig:temp} shows a series of angular distribution templates for different
values of $\alpha_2$ and $p_T^W<\nunit{10}{GeV}$.

\subsubsection{The Treatment of $\alpha_1$}
Since there is no magnetic field in the central charged particle 
tracking detector, it is not possible to identify the charge of the electron.
Without charge identification, this analysis can only be
performed by summing over the $W$ boson charge and polarization.
This implies that the linear term in $\cos \theta^*$ averages
to zero in the limit of complete acceptance.
However, after acceptance cuts have been applied, even the charge averaged
angular distribution does depend on the linear term.
The reason is that events generated with a non-zero  
$\alpha_1$ correspond to slightly more central electrons after they 
are boosted into the lab frame compared to events generated with 
$\alpha_1$ set to zero. After acceptance cuts have been applied, 
fewer events are lost at large $\cos\theta^*$.
However, since this is only 
a second order effect, this measurement is not 
sensitive to  $\alpha_{1}$.
For this analysis, we calculate $\alpha_{1}$ \cite{mi} based on the 
measured \ptw of each event. 
Possible variations of $\alpha_1$ are treated as a source of systematic
uncertainty (see Sec.~\ref{sec:syst}).

\subsection {Backgrounds}
To extract the electron angular distribution from
the transverse mass distribution, the size of the backgrounds has to 
be estimated.
The backgrounds are estimated as functions of the $W$ boson transverse 
momentum and transverse mass, these being the two variables used to extract
the angular distribution. The following sections describe how the four 
dominant backgrounds are calculated, and how they depend on 
transverse mass and transverse momentum. 
\subsubsection{QCD}
\label{QCDBKG}
A large potential source of background is due to 
QCD dijet events, where one
jet is misidentified as an electron and the energy in the event is mismeasured 
resulting in large \met.
This background is estimated using QCD multijet events from our data
following the procedure
described in detail in Ref.~\cite{wzcsprd}. Briefly, the fraction of QCD background
events in the $W$ boson sample is given by
\begin{equation}
  f_{QCD}^{W}=\frac{\epsilon_j}{N_t}\left(\frac{\epsilon_s N_\ell - N_t}
                 {\epsilon_s - \epsilon_j}\right)
\end{equation}
with the following variables:
$N_\ell$ and $N_t$ are the number of events in the $W$ sample
satisfying loose and tight
electron criteria, respectively.
The tight electron  efficiency, $\epsilon_s$, is the fraction of 
loose electrons passing tight cuts as found in a sample of $Z$ boson events,
where one electron is required to pass tight electron identification cuts
and the  other serves as an unbiased probe for determining relative 
efficiencies. 
The jet efficiency, $\epsilon_j$, is the fraction of loose ``fake'' electrons 
that pass tight electron cuts in a sample of multijet events. 
This sample is required 
to have low \met ($<\nunit{15}{GeV}$) 
to minimize the number of $W$ bosons in the sample.
From this analysis,
the overall QCD background fraction is found to be 
$f_{QCD}^W=(0.77 \pm 0.6)\%\:$ with a transverse mass cut of 
$50 < M_T^W < 90$ GeV  imposed, this being the range used in the Bayesian 
analysis. For $f_{QCD}^W$ as a function of $p_T^W$, 
see Table~\ref{tab:bkgfrac}.

\subsubsection{$Z\rightarrow ee$}
Another source of background is 
$Z$ boson events in which one electron is lost in a region of the 
detector that is uninstrumented or one that has a lower 
electron finding efficiency such as that between the CC and the EC. 
This results in
a momentum imbalance, with the event now being indistinguishable from a $W$ 
boson event. This background
can only be estimated using Monte Carlo $Z$ boson events. The number of such $Z$ boson events 
present in the $W$ boson sample is calculated by applying the $W$ boson 
selection cuts to 
{\sc herwig}~\cite{herwig} 
$Z\rightarrow ee$ events that are processed through a 
{\sc geant}~\cite{geant} based 
simulation of the \D0 detector and then overlaid with 
events from random 
$p$\pbar\ crossings. This is done to simulate the underlying event, so that
the effect of the luminosity can be included.
The overall background fraction is found to be
$f_{Z}^{W} = (0.50 \pm 0.06)\% $ averaged over all \ptw.
For the background fraction in each \ptw\ bin, see Table~\ref{tab:bkgfrac}.
\subsubsection{\ttbar Production}
The top quark background is not expected to contribute significantly, 
except in the highest $p_T^W$ bin. The background from these events 
comes from 
$t$ quarks decaying to $W$ bosons. 
If one $W$ boson decays electronically while the other decays into two 
hadronic jets,
the event can mimic a high $p_T$ $W$ boson event.
This background, like the $Z$ boson background, 
is calculated from Monte Carlo using 
{\sc herwig} $t$\tbar events.
The overall background fraction is 
$f_{t\overline{t}}^{W} = (0.087 \pm 0.027)\% $. 
For the background fraction in each \ptw\ bin, see Table~\ref{tab:bkgfrac}.
\subsubsection{$W\rightarrow\tau\nu$}
\begin{figure}[!htbp]
\vbox{
%\vspace{-0.6in}
\centerline{
\epsfig{figure=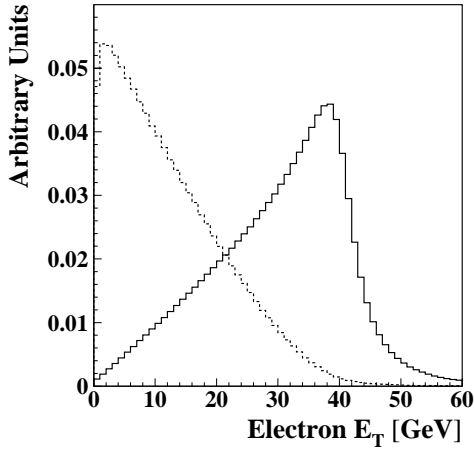,width=7cm}
}
%\vspace{-.5cm}
\caption{Electron $E_T$ spectrum for 
Monte Carlo $W\rightarrow \tau \nu \rightarrow e \nu \nu \nu$ events
(dashed) and $W\rightarrow e\nu$ events (solid histogram). 
Both spectra are normalized to unity for shape comparison. 
\label{fig:tauet}}
}
\end{figure}

$ W\rightarrow \tau \nu$ events in which the $\tau$ decays into an electron 
and two 
neutrinos are indistinguishable from $ W\rightarrow e\nu$ events.
This background is estimated from Monte Carlo simulations 
using the $W$ boson mass  Monte Carlo described above.
A fraction of the events is generated as $W\rightarrow \tau \nu$, decayed 
electronically, with
acceptance and fiducial cuts applied to the decay electron in the 
same manner as in $W\rightarrow e \nu$ events.
The acceptance for $W\rightarrow\tau\nu \rightarrow e \nu \nu \nu$ is 
reduced by 
the branching fraction 
$B(\tau \rightarrow e \nu\nu) = (17.81 \pm 0.07) \%$~\cite{pdg}.
The kinematic acceptance is further reduced by the $E_T$ cut on the 
electron since the three-body decay of the $\tau$ leads to a very soft
electron $E_T$ spectrum  compared to that from 
$W \rightarrow e \nu$ events (see Fig.~\ref{fig:tauet}).
The fraction of $W\rightarrow\tau\nu \rightarrow e \nu \nu \nu$ events
after these cuts are applied to the Monte Carlo 
is $f^W_\tau=(2.03 \pm 0.19)\%$ over all $p^W_T$.

\begin{figure}[!htbp]
\vbox{
%\vspace{-0.2in}
\centerline{
\epsfig{figure=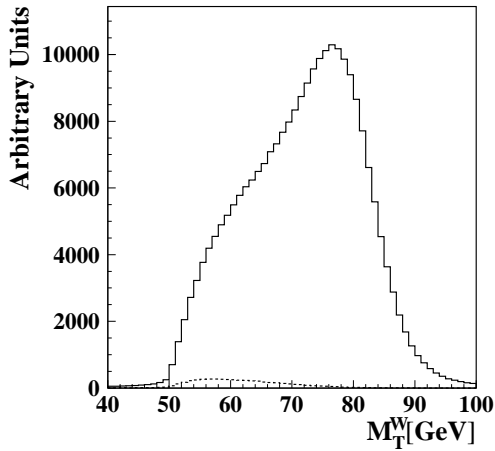,width=7cm}
}
%\vspace{-0.cm}
\caption{Transverse mass distribution for $W \rightarrow e\nu$ events (solid) 
and $W\rightarrow\tau\nu\rightarrow e\nu\nu\nu \: $ events (dashed)  from 
Monte Carlo.\label{fig:taubkg}}
}
\end{figure}

For this analysis, the angular ($\cos\theta^*$) templates are generated
using the $W$ boson mass Monte Carlo simulator 
with the branching ratio 
$B(W \rightarrow \tau \nu) = B(W \rightarrow e \nu)$, assuming lepton 
universality,  
and the above value for $B(\tau \rightarrow e \nu \nu)$. 
The transverse mass of $W\rightarrow\tau\nu$ events 
(Fig.~\ref{fig:taubkg}) is on average lower than that of
$W\rightarrow e\nu$ events, due to the three-body decay of the $\tau$.
\subsubsection{Summary of Backgrounds}
\begin{figure}[!htbp]
\vbox{
\centerline{
\epsfig{figure=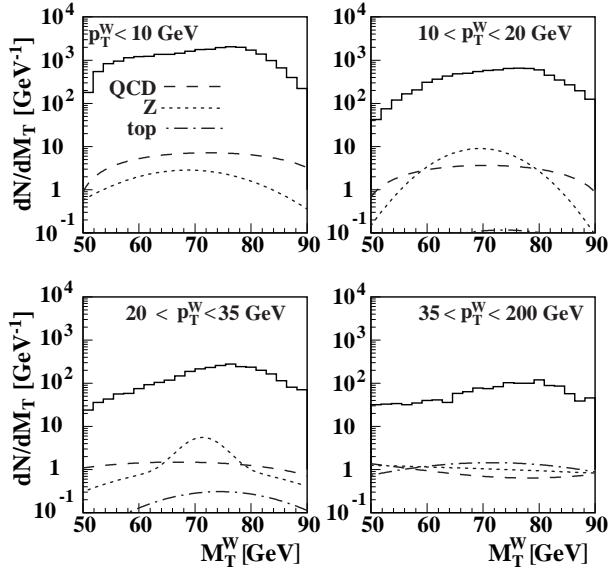,width=10cm}
}
\caption{Transverse mass spectrum for $W \rightarrow e\nu$ candidate events 
(solid histogram) and QCD (dashed), $Z$ boson (dotted), and $t$\tbar\ 
backgrounds 
(dashed-dotted) in four \ptw\ bins.\label{fig:allbkg}}
}
\end{figure}

As we have shown in the previous sections, 
and as can be clearly seen in Fig.~\ref{fig:allbkg},
the background fractions in this measurement are small (a few per cent) over all
$M_T^W$ and $p_T^W$ ranges. The dominant backgrounds are due 
to QCD multijet events and $Z$ boson decays, except in the highest
$p_T^W$ bin where the $t\bar{t}$ background
is comparable in size.
\begin{table}[!htbp]
%\centerline{
\begin{tabular}{l|l|l|l}
\ptw\ [GeV]    & $f_{QCD}^W\;[\%]$ &$f_{Z}^W\;[\%]$ &$f_{t\overline{t}}^W\;[\%]$  \\ \hline
0--10   & $0.6\pm 1.0$ &$0.16 \pm 0.02$ &$0.0028\pm 0.0009$ \\
10--20  & $1.0\pm 1.0$ &$1.1\pm 0.1 $&$0.025 \pm 0.008$ \\
20--35  & $1.3\pm 1.0$ &$1.4\pm 0.2 $&$0.15  \pm 0.05$ \\ 
35--200 & $2.0\pm 1.1$ &$1.7\pm 0.2 $&$2.0   \pm 0.6$ \\ 
\end{tabular}
%}
%\vskip+0.cm
\caption{Background fractions as a function of $p_T^W$ for events with 
a transverse mass cut of 
$50 < M_T^W < 90$ GeV  imposed.\label{tab:bkgfrac}}
\end{table}
\subsection {The Measurement of $\alpha_{2}$} 
\begin{figure}[!htbp]
\vbox{
%\vspace{-0.40in}
%
\centerline{
\epsfig{figure=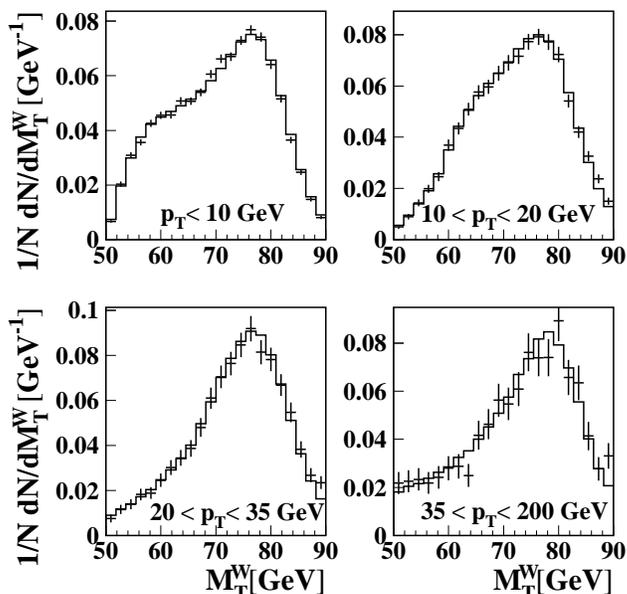,width=9.0cm}}
%\vspace{-0.0in}
\caption{Background subtracted transverse mass distributions (crosses) in 
four \ptw\ bins
compared to Monte Carlo predictions (solid histograms).}
\label{fig:fig8}
}
\end{figure}

\begin{figure}[!htbp]
\vbox{
\centerline{
\epsfig{figure=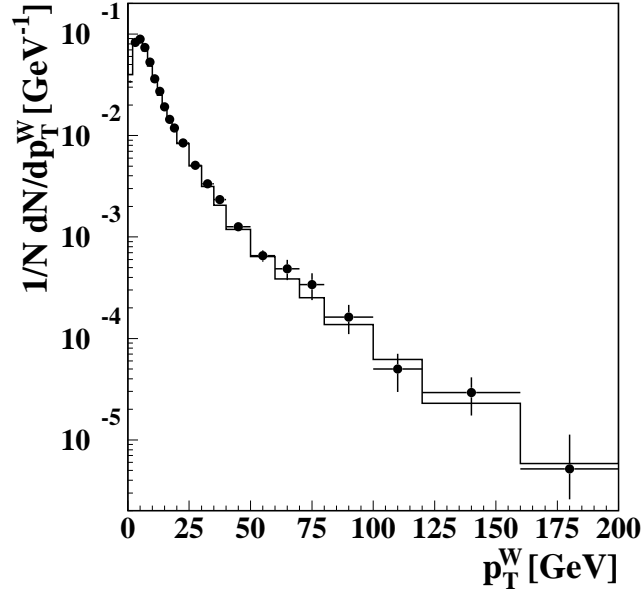,width=9cm}}
\caption{Background subtracted transverse momentum distribution (crosses) 
compared to Monte Carlo prediction (solid histogram). 
The error bars indicate statistical uncertainties only.\label{fig:ptw}}
}
\end{figure}

To obtain the angular distribution for $W$ boson events from data, 
the transverse 
mass distribution is inverted through the use of Bayes' Theorem as 
described in  Sec.~\ref{extraction}. Since the probability distribution 
function
used to invert the $M_T^W$ distribution is generated from Monte Carlo, we 
compare the
background-subtracted $M_T^W$ distribution from data
to that generated through our Monte Carlo to verify that it models the
physics and detector correctly
(see Fig.~\ref{fig:fig8}). Based on a $\chi^2$ test, the agreement between 
data and 
Monte Carlo is good; 
the $\chi^{2}$-probabilities are $11.2\%$,  $80.6\%$,  $93.7\%$ and  $53.7\%$ 
in order of increasing \ptw\ bins.
Likewise, the 
experimental
and Monte Carlo $p_T^W$ distributions can be compared, with the two showing 
agreement with a $\chi^2$-probability of $7.4\%$, where only statistical 
errors are taken into account (see Fig.~\ref{fig:ptw}).

\begin{figure}[!htbp]
\vbox{
%\vspace{-0.2in}
%
\centerline{
\epsfig{figure=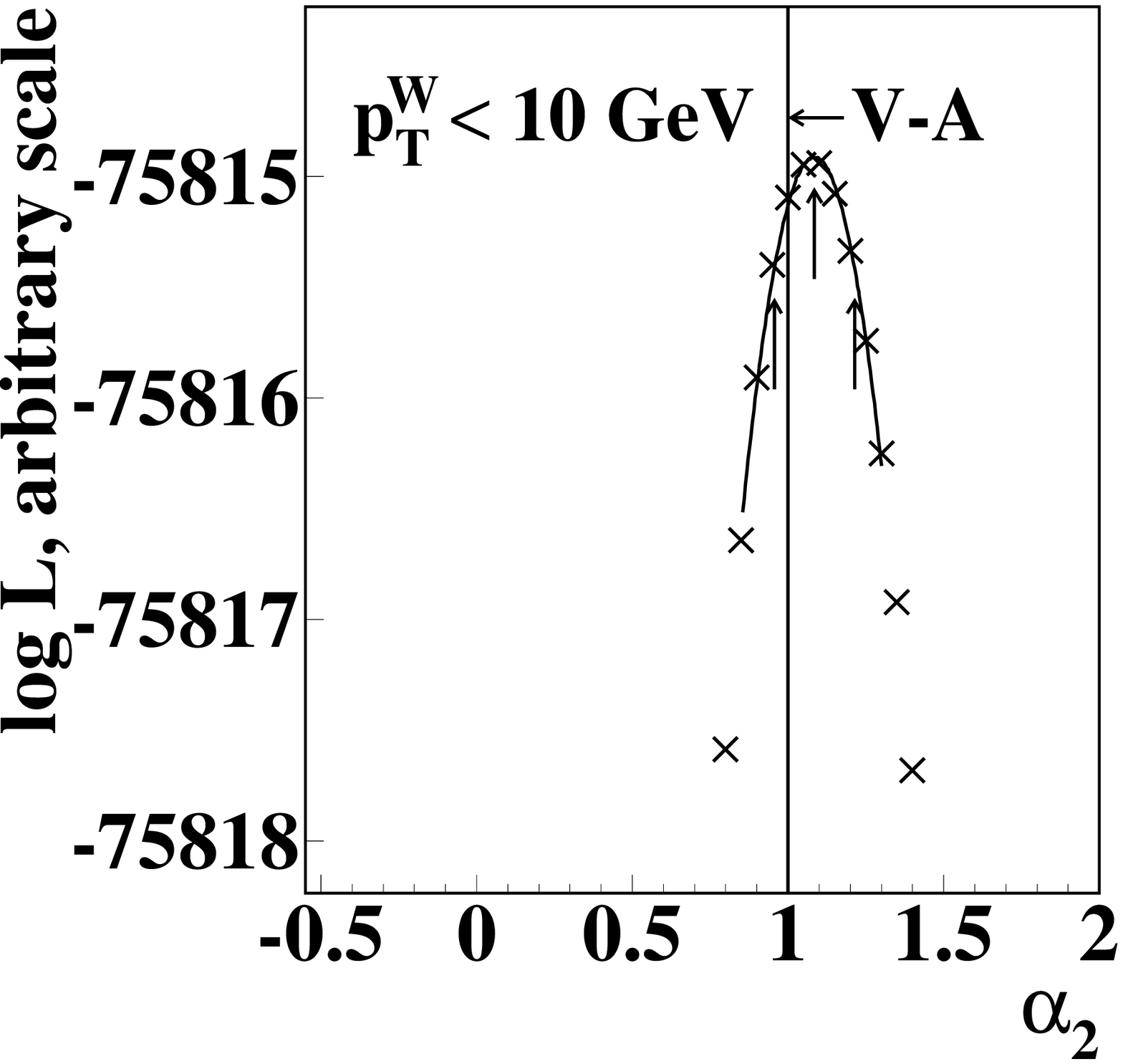,width=4.5cm}
\hspace{-0.5cm}
\epsfig{figure=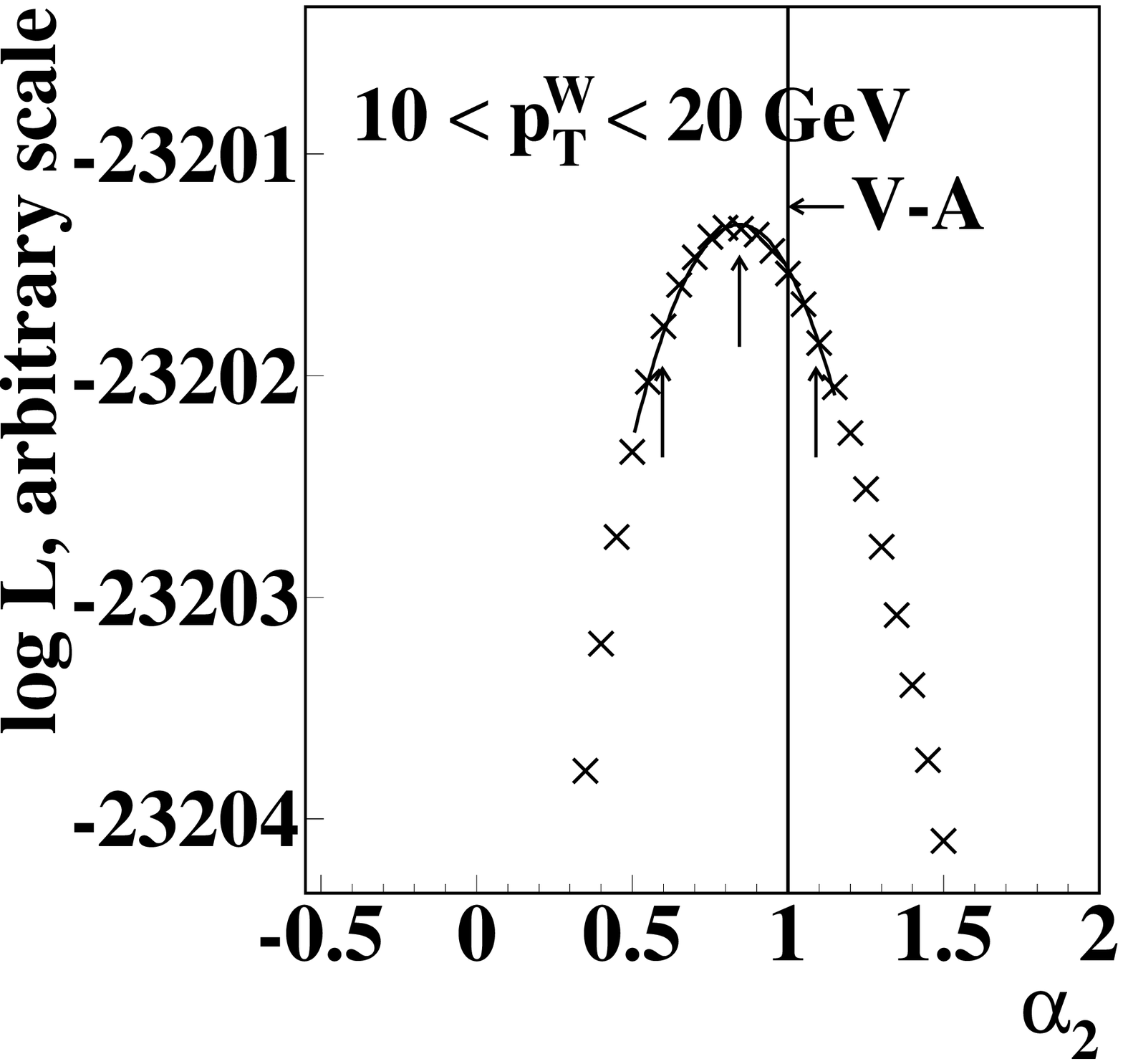,width=4.5cm}
}
\centerline{
\epsfig{figure=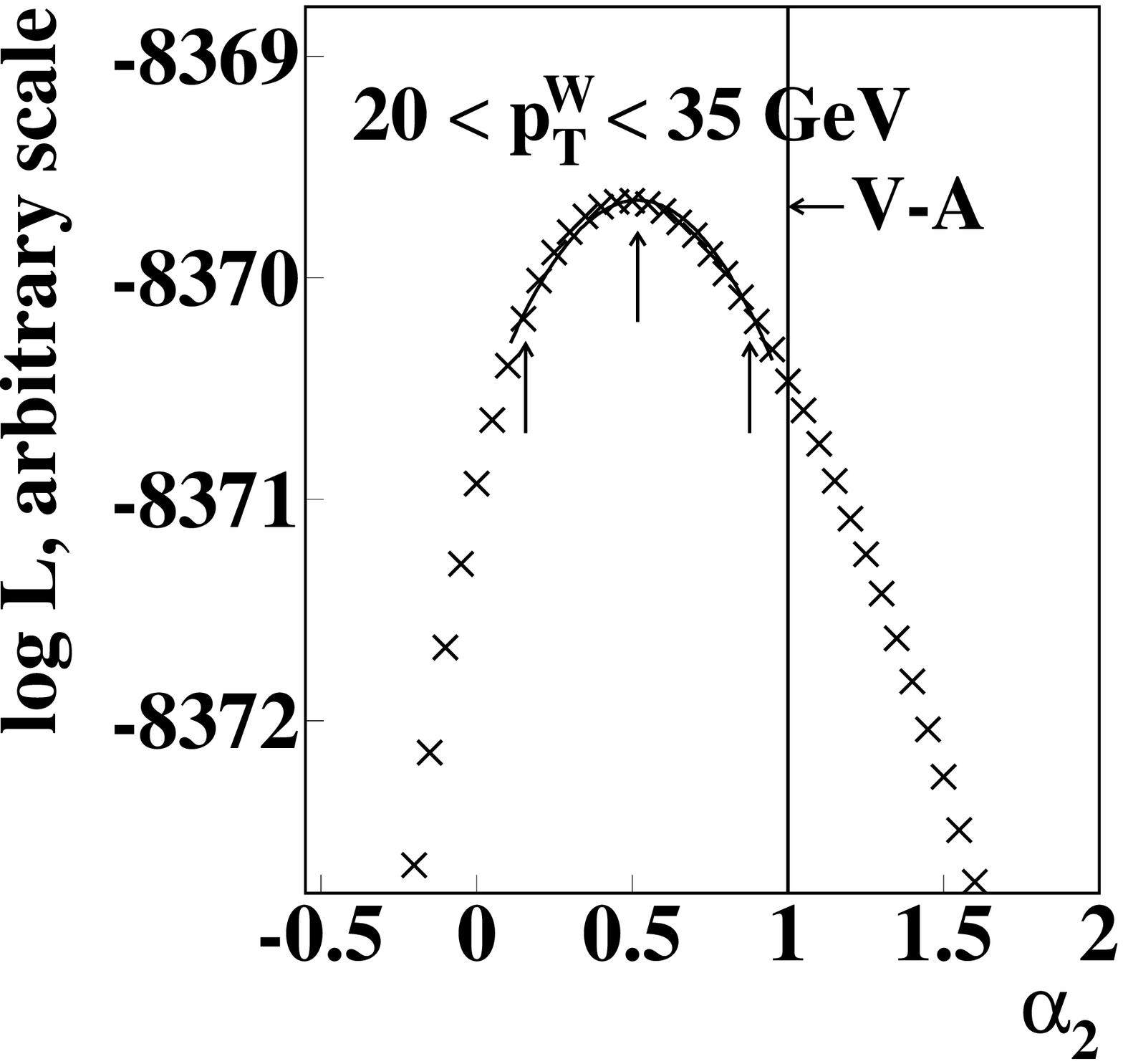,width=4.5cm}
\hspace{-0.5cm}
\epsfig{figure=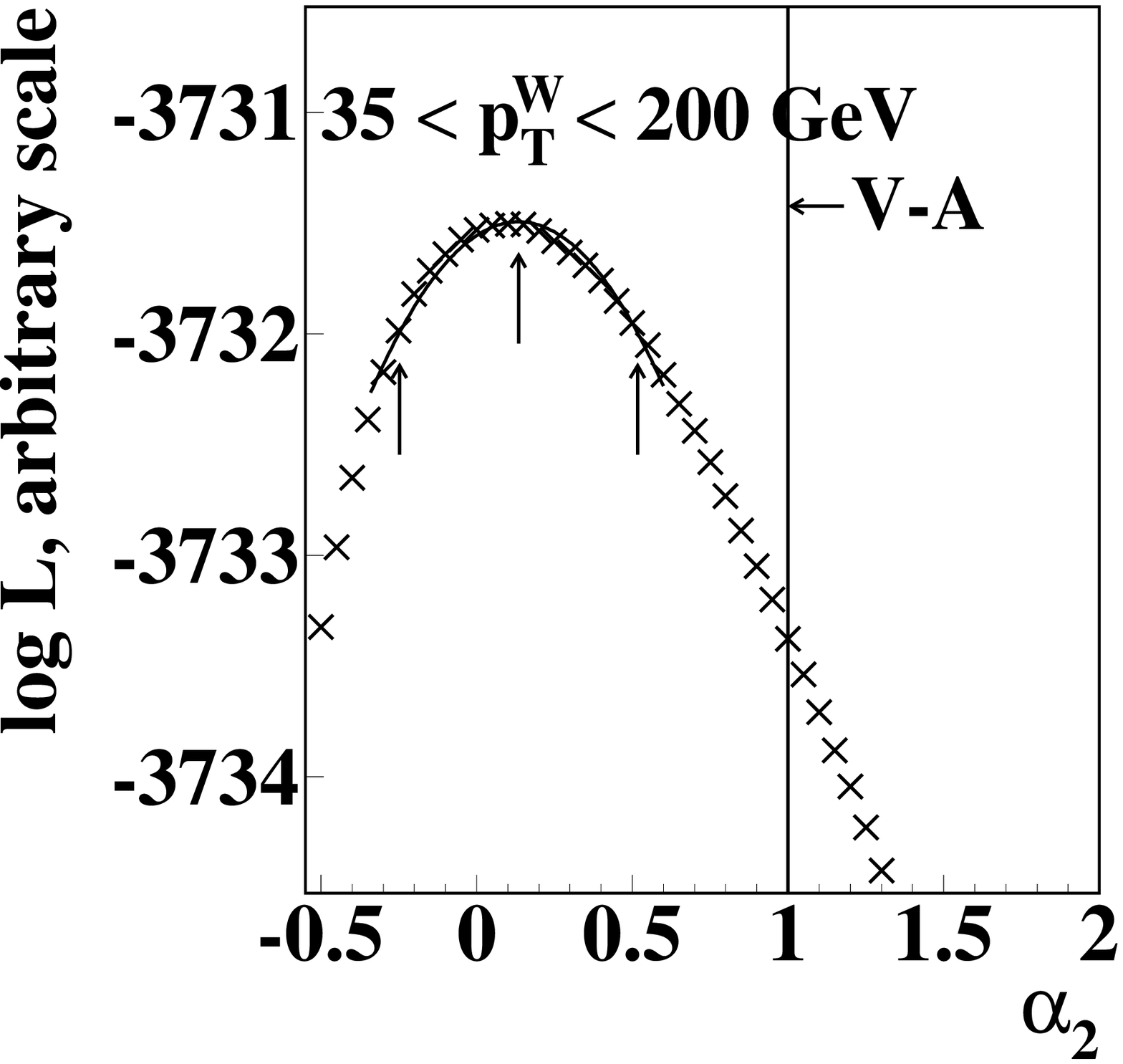,width=4.5cm}
}
%\vspace{-0.5cm}
\caption{Log-likelihood functions for four different \ptw\ bins.
The arrows denote the values of maximum likelihood and the $1\sigma$ errors. 
The vertical lines labeled $V-A$ show $\alpha_1 = 1$, the value for 
$V-A$ theory without QCD corrections.}
\label{fig:loglike}
}
\end{figure}

\begin{figure}[htbp]
\vbox{
%\vspace{-0.2in}
\centerline{
\hspace {0.3cm}
\epsfig{figure=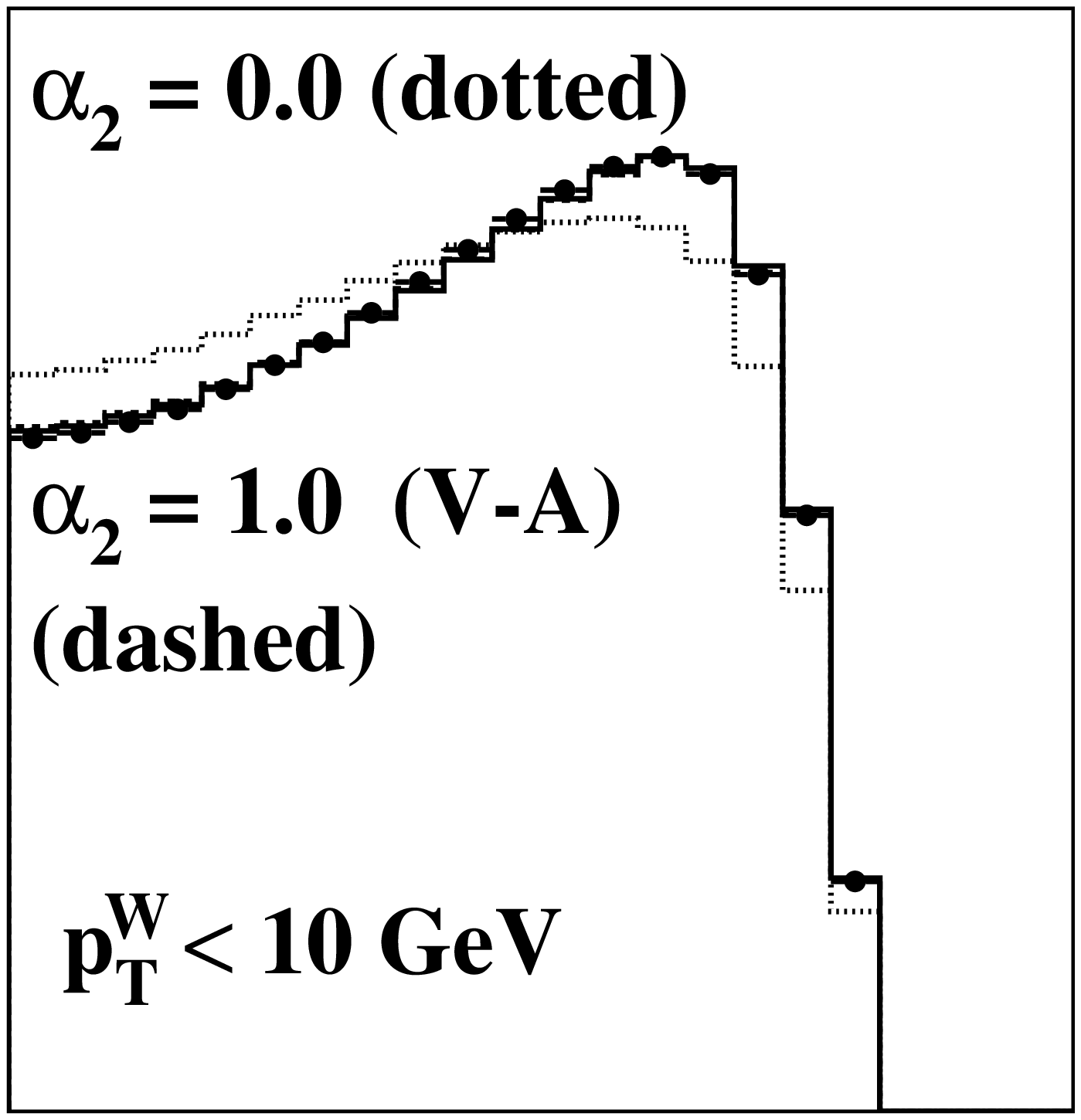,width=5.3cm}
\hspace{-1.7cm}
\epsfig{figure=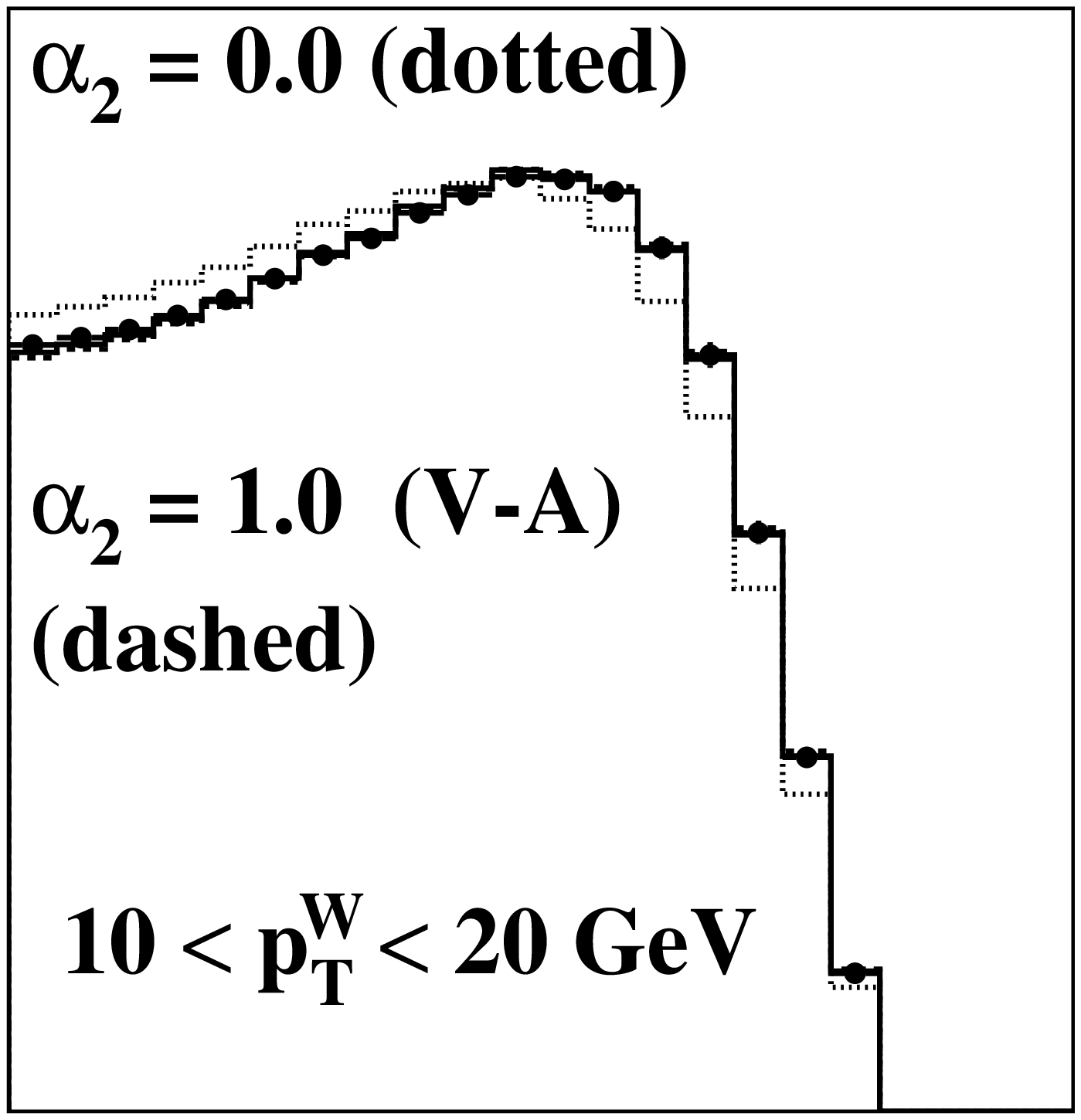,width=5.3cm}
}
\vspace{-1.3cm}
\centerline{
\hspace {0.3cm}
\epsfig{figure=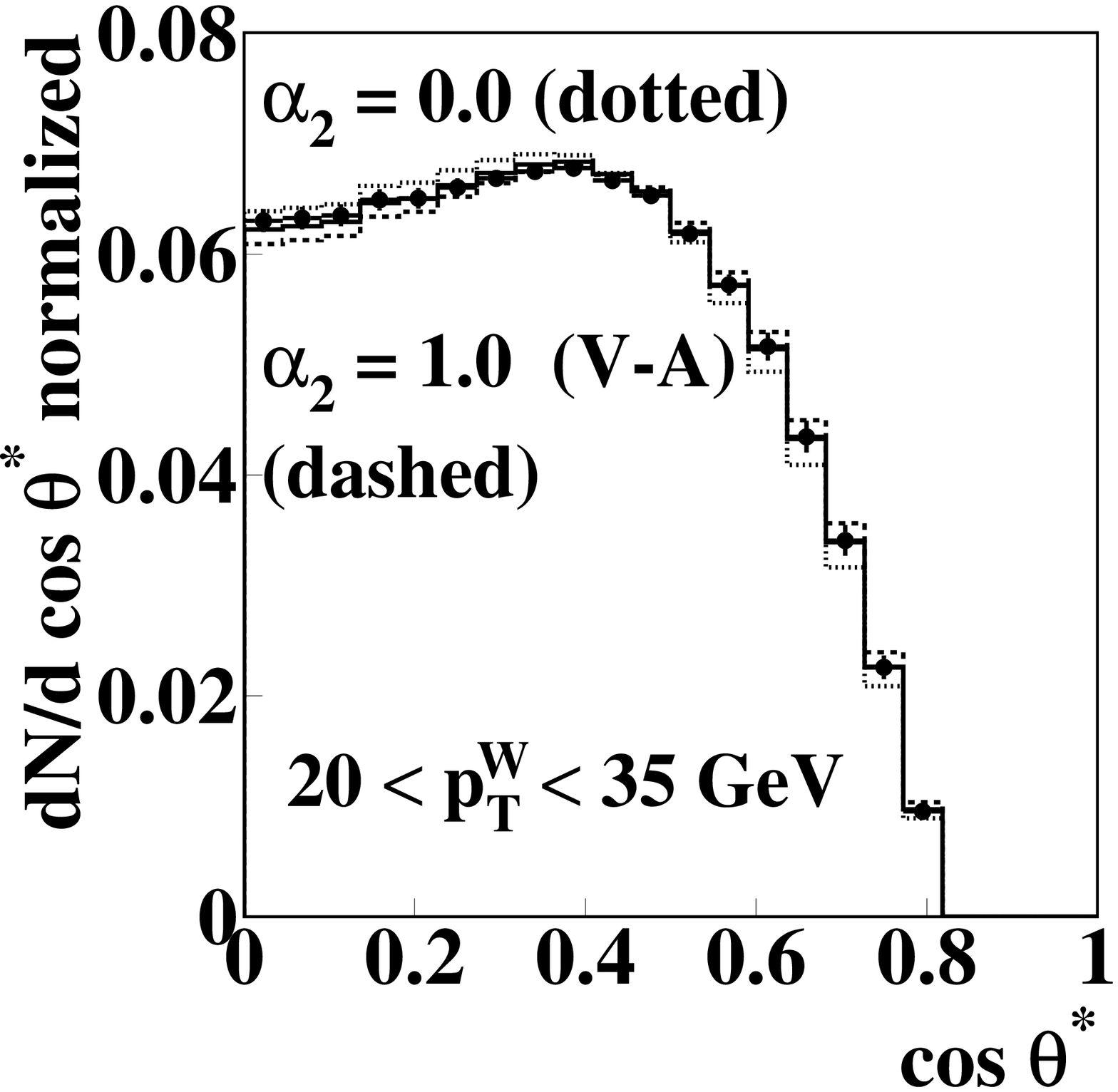,width=5.3cm}
\hspace{-1.7cm}
\epsfig{figure=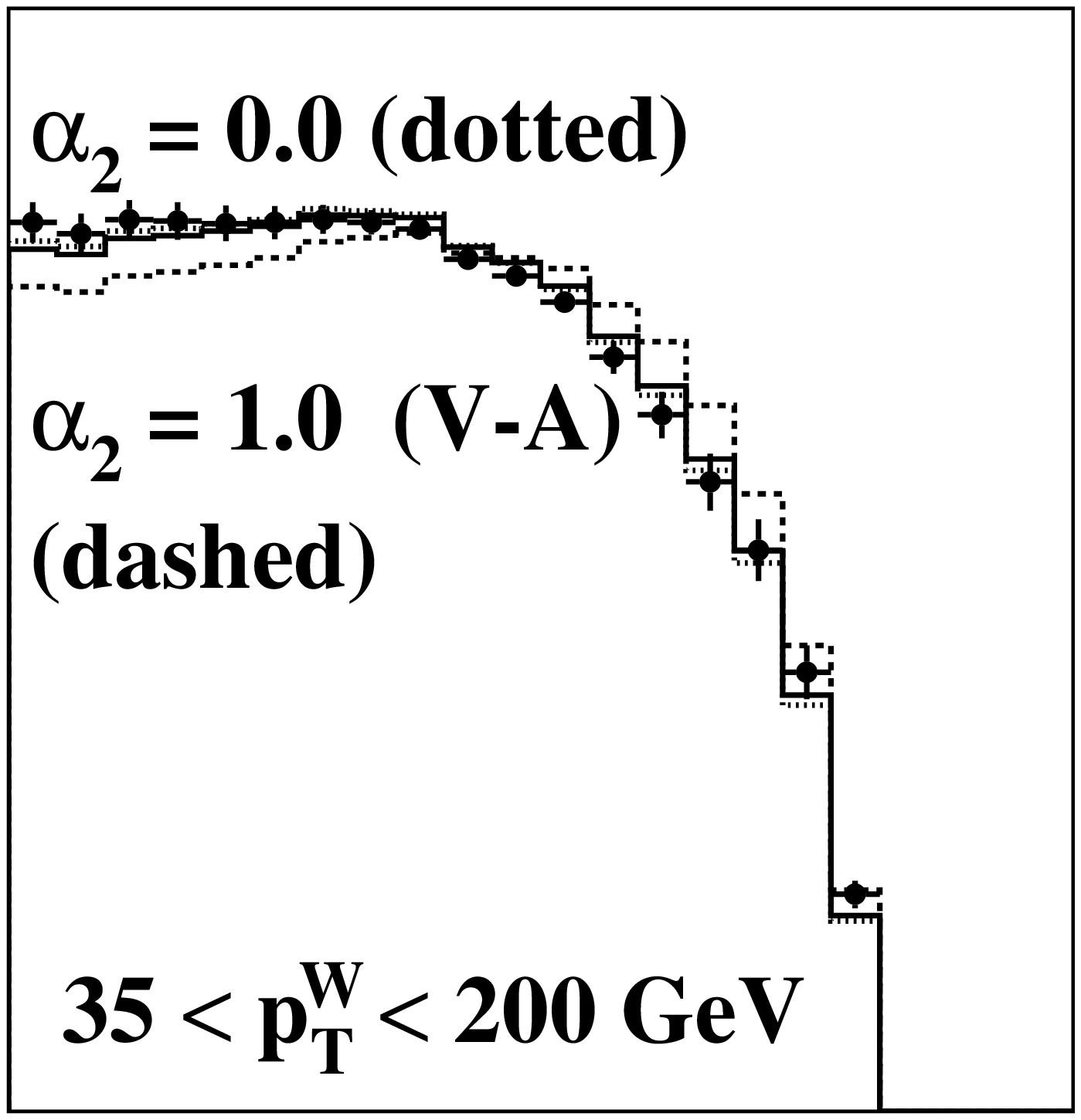,width=5.3cm}
}
\vspace{0.2in}
\caption{Angular distributions for data compared to Monte Carlo 
templates for four different \ptw\ bins. Shown are the templates 
that fit best (solid) and the 
templates for $\alpha_2$ = 1 (dashed) and  
$\alpha_2$ = 0 (dotted).
\label{fig:compang}}
}
\end{figure}

After extracting the angular distribution, the 
parameter $\alpha_{2}$ is computed using the method of maximum 
likelihood (see Fig.~\ref{fig:loglike}).
The angular distribution is compared to a series of Monte Carlo generated 
templates,
each with a different value of $\alpha_2$. The template that results in 
the maximum likelihood gives the value of $\alpha_2$ for each $p_T^W$
bin (Fig.~\ref{fig:compang}). 
The $1 \sigma$ uncertainties in $\alpha_2$ are  
approximately given by the points where the log-likelihood drops by 0.5 units.
To estimate the goodness of fit, the
measured angular distributions are compared to 
these templates using a $\chi ^2$ test. 
The $\chi^2$-probabilities that we obtain are 
$8.4\%$, $59.1\%$, $87.7\%$ and $11.6\%$
in order of increasing \ptw\ bins.
%
%\nopagebreak
%
%\vskip -2.0cm
%
%\clearpage
\subsection{Systematic Errors}
\label{sec:syst}
Systematic errors on our measurement of $\alpha_2$ are due to uncertainties 
in the backgrounds and the parameters used to model the detector in the 
Monte Carlo.
To estimate the errors 
due to the background uncertainties, the parameters from fits of the 
transverse mass
distributions of the background are varied within their errors, and the 
analysis is repeated.
For the errors due to detector modeling,  the corresponding Monte Carlo 
parameters are varied within their errors and the analysis is repeated 
with new angular templates. 
For this analysis, we fixed $\alpha_1$ to the values given by the 
next-to-leading order QCD prediction (see Fig.~\ref{fig:alpha12}). 
The error associated with this choice 
is estimated by changing $\alpha_1$ to the value calculated in the absence 
of QCD effects ($\alpha_1 = 2.0$). 

The dominant systematic errors are due to uncertainties in the 
electromagnetic energy scale and the QCD background.
All systematic errors are summarized in Table~\ref{tab:results}. 
The systematic errors are combined in quadrature. The statistical 
uncertainties are, except for the first \ptw\ bin, larger by a factor of three 
than the systematic uncertainties.
\begin{table}[htbp]
\begin{tabular}{l|l|l|l|l}
\ptw\ [GeV]     &   $0-10$ &   $10-20$ &   
$20-35$  &   $35-200$ \\ \hline
$\alpha_2,\:measured$ &    $1.09$   & $0.84$  & $0.52$  
& $0.13$    \\ 
stat. errors&    $\pm0.13$   & $\pm0.25 $  & $\pm0.36 $  
& $\pm0.38 $    \\ 
$\alpha_2,\: predicted$&    $0.98$   & $ 0.89$  & $ 0.68$  & $ 0.24 $ \\ 
mean \ptw\ &    5.3    &13.3  &25.7  & 52.9  \\ \hline 
QCD & $\pm0.04$ & $\pm0.05$ & $\pm0.09$&  $\pm0.07$ \\  
$Z\rightarrow ee$&    $\pm0.01$ & $\pm0.02$ & $\pm0.02$&  $\pm0.04$ \\  
\ttbar&     $\pm0.00$ & $\pm0.00$ & $\pm0.00$&  $\pm0.02$ \\
EM scale   & $\pm0.06$  & $\pm0.05$ & $\pm0.03$&  $\pm0.04$  \\
hadronic scale  & $\pm0.03$  & $\pm0.01$ & $\pm0.04$&  $\pm0.04$   \\ 
hadronic resol. &     $\pm0.02$ & $\pm0.02$ & $\pm0.05$&  $\pm0.06$ \\ 
fixed $\alpha_1$   & $\pm0.01$ & $\pm0.05$ & $\pm0.03$&  $\pm0.03$ \\ \hline
combined syst. & $\pm0.08$ & $\pm0.09$ & $\pm0.12$&  $\pm0.12$
\end{tabular}
\caption{Central values for $\alpha_{2}$ with statistical and systematic 
errors.\label{tab:results}}
\end{table}
\subsection{Results and Sensitivity}
\begin{figure}[htbp]
\vbox{
%\vspace{-0.2in}
\centerline{
\epsfig{figure=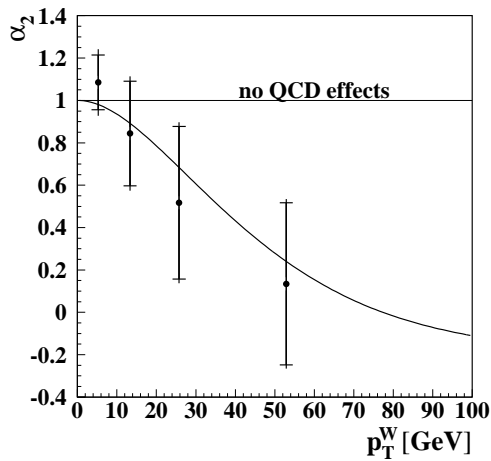,height=7cm}
}
\caption{Measured $\alpha_{2}$ as a function of \ptw\ compared 
to the next-to-leading order QCD calculation by Mirkes (curve) 
and calculation in the absence of QCD (horizontal line).
The combined systematic and statistical errors are shown as vertical bars, 
while the statistical errors alone are marked by horizontal ticks. 
}
\label{fig:result}
}
\end{figure}

To estimate the sensitivity of this experiment, the $\chi^{2}\:$ of the 
$\alpha_{2}$ distribution is calculated 
with respect to the prediction of the $V-A$
theory modified by next-to-leading order QCD and that of the $V-A$ theory 
in the absence of QCD corrections.  
The  $\chi^{2}\:$  with respect to the QCD prediction is 0.8 for 4 degrees of
freedom,  which corresponds to a probability of $94\%$.
The  $\chi^{2}\:$ with respect to pure $V-A$ is 7.0 for 4 degrees of
freedom,  which 
corresponds to $14 \%$ probability. 
To make a more quantitative estimate of how much
better $V-A$ modified by next-to-leading order QCD agrees over pure $V-A$,
we use the odds-ratio method
\footnote{
The odds-ratio $R$ is defined as 
$R=\frac{\prod_{i}p_{i}(\alpha_{2}({\rm NLO \; QCD}))}{\prod_{i}p_{i}(\alpha_{2}({\rm no\; QCD}))}$
where the product is over \ptw\ bins, 
$p_{i}(\alpha_{2}({\rm NLO\;QCD}))\:$is the normalized probability at the 
predicted value for
$\alpha_{2}\:$ for the $i^{th}$ \ptw\ bin, $p_{i}(\alpha_{2}({\rm no\; QCD}))\:$ 
is the normalized probability at the 
predicted value for $V-A$ theory without QCD effects, i.e. at $\alpha_{2}=1.0$.
This corresponds to a $1 \sigma$ separation for $\log(R)=0.5$.
},
which prefers the former over the latter theory
by $\approx 2.3\, \sigma$.
The results
of our measurement along with the theoretical prediction are given in
Fig.~\ref{fig:result} and Table~\ref{tab:results}.
%
%\nopagebreak
%\vskip-0.5cm

\section{Conclusions}
\label{sec:concl}
Using data taken with the \D0 detector during the 1994--1995 Fermilab 
Tevatron collider
run, we have presented a measurement of the 
angular distribution of decay electrons from $W$ boson events.  
A next-to-leading order QCD calculation is preferred by $\approx 2.3 \,\sigma$ 
over a calculation where no QCD effects are included. 

\section{Acknowledgments}

\input{acknowledgement_paragraph.tex}

%-------------------------------------------------------------------------------

\end{document}

%% file: list_of_authors.tex
% LIST_OF_AUTHORS.TEX                 8/10/00            
%
\author{                                                                      
%% names begin here                                                           
B.~Abbott,$^{50}$                                                             
M.~Abolins,$^{47}$                                                            
V.~Abramov,$^{23}$                                                            
B.S.~Acharya,$^{15}$                                                          
D.L.~Adams,$^{57}$                                                            
M.~Adams,$^{34}$                                                              
G.A.~Alves,$^{2}$                                                             
N.~Amos,$^{46}$                                                               
E.W.~Anderson,$^{39}$                                                         
M.M.~Baarmand,$^{52}$
% ...
\\                                                                            
\vskip 0.30cm                                                                 
\centerline{(D\O\ Collaboration)}                                             
\vskip 0.30cm                                                                 
}                                                                             
\address{                                                                     
\centerline{$^{1}$Universidad de Buenos Aires, Buenos Aires, Argentina}       
\centerline{$^{2}$LAFEX, Centro Brasileiro de Pesquisas F{\'\i}sicas,         
                  Rio de Janeiro, Brazil}                                     
\centerline{$^{3}$Universidade do Estado do Rio de Janeiro,                   
                  Rio de Janeiro, Brazil}                                     
\centerline{$^{4}$Institute of High Energy Physics, Beijing,                  
                  People's Republic of China}                                 
\centerline{$^{5}$Universidad de los Andes, Bogot\'{a}, Colombia}             
\centerline{$^{6}$Charles University, Prague, Czech Republic}                 
\centerline{$^{7}$Institute of Physics, Academy of Sciences, Prague,          
                  Czech Republic}                                             
\centerline{$^{8}$Universidad San Francisco de Quito, Quito, Ecuador}         
\centerline{$^{9}$Institut des Sciences Nucl\'eaires, IN2P3-CNRS,             
                  Universite de Grenoble 1, Grenoble, France}                 
\centerline{$^{10}$CPPM, IN2P3-CNRS, Universit\'e de la M\'editerran\'ee,     
                  Marseille, France}                                          
\centerline{$^{11}$LPNHE, Universit\'es Paris VI and VII, IN2P3-CNRS,         
                  Paris, France}                                              
\centerline{$^{12}$DAPNIA/Service de Physique des Particules, CEA, Saclay,    
                  France}                                                     
\centerline{$^{13}$Panjab University, Chandigarh, India}                      
\centerline{$^{14}$Delhi University, Delhi, India}                            
\centerline{$^{15}$Tata Institute of Fundamental Research, Mumbai, India}     
\centerline{$^{16}$Seoul National University, Seoul, Korea}                   
\centerline{$^{17}$CINVESTAV, Mexico City, Mexico}                            
\centerline{$^{18}$FOM-Institute NIKHEF and University of                     
                  Amsterdam/NIKHEF, Amsterdam, The Netherlands}               
\centerline{$^{19}$University of Nijmegen/NIKHEF, Nijmegen, The               
                  Netherlands}                                                
\centerline{$^{20}$Institute of Nuclear Physics, Krak\'ow, Poland}            
\centerline{$^{21}$Institute for Theoretical and Experimental Physics,        
                   Moscow, Russia}                                            
\centerline{$^{22}$Moscow State University, Moscow, Russia}                   
\centerline{$^{23}$Institute for High Energy Physics, Protvino, Russia}       
\centerline{$^{24}$Lancaster University, Lancaster, United Kingdom}           
\centerline{$^{25}$University of Arizona, Tucson, Arizona 85721}              
\centerline{$^{26}$Lawrence Berkeley National Laboratory and University of    
                  California, Berkeley, California 94720}                     
\centerline{$^{27}$University of California, Davis, California 95616}         
\centerline{$^{28}$California State University, Fresno, California 93740}     
\centerline{$^{29}$University of California, Irvine, California 92697}        
\centerline{$^{30}$University of California, Riverside, California 92521}     
\centerline{$^{31}$Florida State University, Tallahassee, Florida 32306}      
\centerline{$^{32}$University of Hawaii, Honolulu, Hawaii 96822}              
\centerline{$^{33}$Fermi National Accelerator Laboratory, Batavia,            
                   Illinois 60510}                                            
\centerline{$^{34}$University of Illinois at Chicago, Chicago,                
                   Illinois 60607}                                            
\centerline{$^{35}$Northern Illinois University, DeKalb, Illinois 60115}      
\centerline{$^{36}$Northwestern University, Evanston, Illinois 60208}         
\centerline{$^{37}$Indiana University, Bloomington, Indiana 47405}            
\centerline{$^{38}$University of Notre Dame, Notre Dame, Indiana 46556}       
\centerline{$^{39}$Iowa State University, Ames, Iowa 50011}                   
\centerline{$^{40}$University of Kansas, Lawrence, Kansas 66045}              
\centerline{$^{41}$Kansas State University, Manhattan, Kansas 66506}          
\centerline{$^{42}$Louisiana Tech University, Ruston, Louisiana 71272}        
\centerline{$^{43}$University of Maryland, College Park, Maryland 20742}      
\centerline{$^{44}$Boston University, Boston, Massachusetts 02215}            
\centerline{$^{45}$Northeastern University, Boston, Massachusetts 02115}      
\centerline{$^{46}$University of Michigan, Ann Arbor, Michigan 48109}         
\centerline{$^{47}$Michigan State University, East Lansing, Michigan 48824}   
\centerline{$^{48}$University of Nebraska, Lincoln, Nebraska 68588}           
\centerline{$^{49}$Columbia University, New York, New York 10027}             
\centerline{$^{50}$New York University, New York, New York 10003}             
\centerline{$^{51}$University of Rochester, Rochester, New York 14627}        
\centerline{$^{52}$State University of New York, Stony Brook,                 
                   New York 11794}                                            
\centerline{$^{53}$Brookhaven National Laboratory, Upton, New York 11973}     
\centerline{$^{54}$Langston University, Langston, Oklahoma 73050}             
\centerline{$^{55}$University of Oklahoma, Norman, Oklahoma 73019}            
\centerline{$^{56}$Brown University, Providence, Rhode Island 02912}          
\centerline{$^{57}$University of Texas, Arlington, Texas 76019}               
\centerline{$^{58}$Texas A\&M University, College Station, Texas 77843}       
\centerline{$^{59}$Rice University, Houston, Texas 77005}                     
\centerline{$^{60}$University of Washington, Seattle, Washington 98195}       
}                                                                             
%end                                                                          

%% file: acknowledgement_paragraph.tex
% Acknowledgement_paragraph.tex
%
We thank the staffs at Fermilab and at collaborating institutions 
for contributions to this work, and acknowledge support from the 
Department of Energy and National Science Foundation (USA),  
Commissariat  \` a L'Energie Atomique and
CNRS/Institut National de Physique Nucl\'eaire et 
de Physique des Particules (France), 
Ministry for Science and Technology and Ministry for Atomic 
   Energy (Russia),
CAPES and CNPq (Brazil),
Departments of Atomic Energy and Science and Education (India),
Colciencias (Colombia),
CONACyT (Mexico),
Ministry of Education and KOSEF (Korea),
CONICET and UBACyT (Argentina),
A.P. Sloan Foundation,
and the A. von Humboldt Foundation.
%